\documentstyle[preprint,aps]{revtex}
\draft
\begin{document}
\preprint{P-95-01-005}
\draft
\title{
Kinetics of phase ordering with topological textures}
\author{Martin Zapotocky \cite{MZ}}
\address{
Department of Physics and Materials Research Laboratory,\\
University of Illinois at Urbana-Champaign,
Urbana, Illinois 61801, USA}
\author{Wojtek Zakrzewski \cite{WZ}}
\address{Department of Mathematical Sciences, University of Durham,
Durham DH1 3LE, UK}
\date{January 30, 1995}
\maketitle
\begin{abstract}
We study the role played by topological textures and antitextures
during the phase ordering of a two-dimensional system described by the
discretised nonlinear O($3$) sigma model with purely dissipative dynamics.  We
identify and characterise two distinct mechanisms for the decay of the
order parameter variations --- single texture unwinding,
and topological charge annihilation.  Our results show that while at
early times after the quench, the annihilation process dominates, the
unwinding processes become of comparable importance at later times. We
calculate the correlations in the order parameter and in the
topological charge density, and show that dynamical scaling is
strongly violated due to the occurence of multiple length scales
growing differently in time.
\end{abstract}
\pacs{PACS numbers: 64.60Cn, 82.20Mj, 05.70.Fh, 11.27.+d}
\vfil\eject
The field of phase ordering kinetics, investigating the time evolution
of a system quenched from the disordered phase into the ordered phase, has
attracted considerable attention in recent years \cite{Bray-review}.
It has been shown that many features of phase ordering in systems supporting
topologically stable singular defects (for example, in systems described by the
O($N$) vector model in $d$ dimensions with $d \ge N$ \cite{vector}, or
in two and three-dimensional nematic liquid crystals \cite{nematic})
can be understood theoretically by investigating the dynamics of the
numerous topological defects generated during the quench. In systems
where topologically stable singular objects cannot occur (for example,
in the O($N$) model system in dimension $d < N$), such an approach
cannot be used. A special and interesting case is that of the O($N$)
model system in $N-1$ spatial dimensions, which supports topologically
stable, but non-singular objects -- topological textures. The purpose
of this communication is to report results from an investigation of the
role played by topological textures during the phase ordering of
an O($3$) vector model system in two
spatial dimensions. This system has been
previously investigated by Toyoki \cite{Toyoki}, who calculated the
time dependence of the order parameter correlation function, and by
Bray and Humayun \cite{Bray+Humayun}, who investigated the decay of
the free energy in the system. A detailed analysis of the phase
ordering process in terms of the behavior of the topological objects
present in this system, however, has not been previously given
\cite{Rutenberg}.

The model investigated in our simulations is the non-linear O($3$)
sigma-model on a two-dimensional lattice. The order parameter $\bf m$
is correspondingly a 3-component vector with unit magnitude;
the local value of the order parameter ${\bf m}({\bf r},t)$ will be refered to
as the spin.
The phase-ordering simulation is started with randomly oriented spins,
corresponding to the configuration
generated immediately after a quench to zero temperature. The
configuration is then evolved using the
dissipative dynamical equation
\begin{equation}
{\partial {\bf m} \over \partial t} = \nabla^2 {\bf m} - ({\bf m}
\cdot \nabla^2{\bf m}) {\bf m} \,,
\label{dynamics}
\end{equation}
where the second term on the right hand side enforces the constraint
${\bf m}\cdot {\bf m} = 1$. To adequately describe the phase-ordering
process, Eq.~(\ref{dynamics}) must be regularized on a scale given by
the equilibrium bulk correlation length; we effectively impose the
regularization condition by considering Eq.~(\ref{dynamics}) on a
discrete lattice.  We evolved Eq.~(\ref{dynamics}) using the technique
developed in Ref.~\cite{Wojtek} (based on the 4th order Runge Kutta
method) appropriately adapted to our case.  The spatial discretization
step was taken to be $dx=dy=0.1$, the time step was $dt=0.0002$, and
we worked with periodic boundary conditions.  We used system sizes
between 252 and 512 lattice units, and our longest runs reached times
$t=40$.

We now briefly review the concept of a topological texture
\cite{Rajaraman}. The spin
configuration of a single texture in an infinite continuum system is
given by
\begin{equation}
m_x({\bf r}) = {4 a x \over r^2+4a^2}\ , \
m_y({\bf r}) = {4 a y \over r^2+4a^2}\ , \
m_z({\bf r}) = {r^2-4 a^2 \over r^2+4 a^2}\ .
\label{texture}
\end{equation}
The orientation of the spin changes from up in the center of the
texture ($r=0$) to down at the boundary of the system ($r=\infty$),
going through a vortex-like configuration with spins pointing radially
outwards on the circle $r= 2 a$.  It is easily seen that the spin
configuration (\ref{texture}) covers the order parameter space (given
in our case by the unit sphere in three dimensions) exactly once,
corresponding to a topological charge of one. It is possible to show
that the configuration (\ref{texture}) (or any global rotation
thereof) has the minimum energy (with energy density taken as
${1 \over 8 \pi} ({\bf
\nabla m})^2$) of all configurations with topological charge one.
The value of the minimum total energy is $E=1$. By an {\it antitexture},
we mean a configuration similar to (\ref{texture}), but with $m_y({\bf
r})$ replaced by $-m_y({\bf r})$. This configuration wraps around the
spin sphere once, but in the opposite sense compared to
(\ref{texture}), corresponding to topological charge $-1$.

The crucial quantity for our investigation of the role of textures and
antitextures in the phase ordering process is the {\it topological
charge density} $q({\bf r})$
\begin{equation}
q({\bf r}) = {1 \over 4 \pi} {\bf m}\cdot(\partial_x {\bf m} \times
\partial_y {\bf m})
\label{charge}
\end{equation}
which, when integrated over the whole system, gives the total
topological charge. For the single texture configuration
(\ref{texture}), the topological charge density has the form
$q(r)={1 \over \pi} 4 a^2/ (r^2+4 a^2)^2$, and exhibits a pronounced
peak at $r=0$ with half-width $1.287\cdot a$. For a single
antitexture, $q(r)$ is of the same form, only negative.

In Fig.~(\ref{big}) a-c, we plot the topological charge density at a
progressive series of times in a section of a system undergoing phase
ordering.. The plots exhibit numerous well defined peaks and
antipeaks, corresponding at later times to rather well separated
textures and antitextures of varying sizes. The average separation
between the textures (or antitextures) grows, and at the
latest time, the system is strongly ``intermittent'' in the sense that
the topological charge density differs significantly from zero only
within very well localised regions, with the spin configuration
practically homogenous in between.  Note that the ``typical'' texture
size does not appreciably increase with time.

A detailed inspection of series of plots similar to Fig.~1, taken at
closer values of time, reveals that the variations in topological
charge density decay through two distinct processes: single texture
(or antitexture) unwinding, and topological charge annihilation. The
first process appears as a growing isolated peak in topological charge
density, and corresponds to a localised configuration of type
(\ref{texture}) with decreasing size $a$. Such a process would conserve the
total topological charge and the total energy of the texture in a
continuum system; in a discrete system, however, this conservation is
strongly violated once the texture size decreases to several lattice
spacings. Eventually, the texture comes (up to a global rotation)
close to the extreme configuration where the spin points up at the
center lattice point, and down everywhere else; such a configuration
has only $1 /2 \pi$ of the original texture energy. This
configuration is followed by a flip of the central spin, and the
complete disappearance of the texture \cite{soft_spin}. The size $a$
of the shrinking texture in our simulation varied very roughly as
$\tau ^{1/4}$, where $\tau$ is the time remaining to the flip.

The second process visible in the topological charge density plots is
the mutual annihilation of overlapping regions of positive and
negative topological charge density (overlapping in the sense that
they are not separated by a region where $q({\bf r})=0$). In
Fig.~(\ref{annih}), we show the evolution starting from a slightly
overlapping texture - antitexture pair. The height of
the two peaks decays, and the overlap of the regions of positive and
negative $q({\bf r})$ increases with time. The peaks initially move
slightly together, but later move significantly apart
\cite{annihilation}. It is important to realise that this
``texture-antitexture annihilation'' process differs radically from
the process of annihilation of a singular defect with it's antidefect
({\it e.g.} a vortex and an antivortex in the 2d O($2$) model), where
the singular cores keep their identities and gradually approach.  In
the texture-antitexture annihilation, the total charge enclosed by
each of the regions of positive and negative $q({\bf r})$ gradually
decays to zero, and the annihilation of topological charge occurs
independently of whether a complete texture and antitexture are
present. In contrast to this, the mechanism of unwinding (discussed in
the previous paragraph) occurs only if the unwinding region encloses a
total topological charge close to $1$ or $-1$.

In order to assess the relative importance of the two processes
discussed above during phase ordering, we
investigated the time dependence of the quantities $Q_+$ and $Q_-$,
defined by
\begin{equation}
Q_+=\int d^2r\, \hbox{\rm {max}}[q({\bf r}),0]\,,\ \ \ \ \ Q_-=\int d^2r\,
\hbox{\rm{min}}[q({\bf r}),0]
\label{q+q-}
\end{equation}
where the integral is over the whole system. In a system with well
separated textures and antitextures, $Q_+$ counts the number of
textures, $Q_-$ counts the number of antitextures, $Q=Q_++Q_-$ gives
the total topological charge, and $P=Q_+-Q_-$ counts the total number
of topological objects in the system. Figure (\ref{steps}) shows that
at late times, the the total topological charge $Q=Q_++Q_-$ varies
only in sharp steps of size $-1$ or $+1$, corresponding to an
unwinding of a single texture, resp. antitexture. The steps are also
visible in the corresponding $Q_+(t)$ or $Q_-(t)$ curve
(see Fig.~(\ref{steps})). Note, however, that while the $Q(t)$ curve
is flat, the $Q_+(t)$ and $Q_-(t)$ curves decay significantly in
between the unwinding steps. This demonstrates that the process of
topological charge annihilation, as defined in the previous paragraph,
takes place. The relative importance of the unwinding and annihilation
processes in a given time range is given by the ratio of the number of
steps in the $Q(t)$ curve to the total drop of the integrated absolute
topological charge density $P=Q_+-Q_-$. In the time interval between
$t=3$ and $t=30$, this ratio is $\rho=0.45$, showing that in this time
range, unwindings and the annihilation processes play almost equally
important roles. At earlier times, it is difficult to calculate the
ratio $\rho$, as the total topological charge $Q$ of a large system no
longer exhibits well separated steps. A lower bound for $\rho$,
however, may be still obtained by counting the sharp peaks (see the
inset in Fig.~(\ref{steps})) occuring in the curve $q_{max}(t)$, where
$q_{max}$ is defined as the maximum of $|q({\bf r})|$ over the whole
system.  Each peak corresponds to the final stages of shrinking and
consequent flipping of a texture or antitexture; however, if two
textures unwind at almost the same time in two different parts of the
system, only one peak may be visible. Comparing the number of peaks
with the drop in $P(t)$ in the time interval from $t=0.5$ to $t=1.0$
gives $\rho \ge 0.25$. This is consistent with the expectation that
since the textures and antitextures are better separated as $t$
increases [see Fig.~(\ref{big})], $\rho$ should increase with time.

We now present results averaged over 14 runs in a system of size 512,
evolved until $t=1.6$. Longer runs in systems of size 252 gave similar
results. The total topological charge $Q$ was approximately conserved
and close to zero ($|Q(t)|<5$ at all $t$). In Fig.~(\ref{energy}), we
plot the integrated absolute charge $P=Q_+-Q_-$, the free energy $E$,
and their difference.  The asymptotic equality of $E$ and $P$
indicates that at late times, the system is
well separated into textures and antitextures, each of energy 1. Note
that the inequality $E(t) > P(t)$ is satisfied at all times; this is
consistent with the well known \cite{Belavin} global inequality $E \ge
\vert Q \vert$ being valid inside each region containing a texture or
an antitexture \cite{ineq}.  Both $P$ and $E$ decay asymptotically as
$t^{-0.64 \pm 0.02}$,
indicating that the average separation $D(t)$ between topological
objects (textures or antitextures) grows as $t^{0.32 \pm 0.01}$. Note
that this differs significantly from the dimensional analysis
prediction of a length scale growing as $t^{1/2}$, and points towards
the presence of scaling violations.  In contrast, the difference
$E(t)-P(t)$, characterising topologically trivial spin variations,
decays asymptotically as $t^{-0.92 \pm 0.03}$ [see the inset in
Fig.~(\ref{energy}] which agrees much better with the dimensional
analysis result. The onset of the approximate power-law regime for
$P(t)$ occurs at $t \simeq 0.02$, corresponding to the time after
which well formed textures and antitextures are seen in the
topological charge density plots.

We calculated three separate correlation functions: the spin-spin
correlation $C(r,t)=\langle {\bf m}({\bf x},t)\cdot{\bf m}({\bf x}+{\bf
r},t)\rangle $, the topological charge density correlation
$C_q(r,t)=\langle q({\bf x},t) q({\bf x}+{\bf r},t)\rangle / \langle
q({\bf x},t) q({\bf x},t) \rangle $, and the
correlation of the absolute topological charge density,
$C_p(r,t)=\langle p({\bf x},t) p({\bf x}+{\bf r},t)\rangle / \langle
p({\bf x},t) p({\bf x},t) \rangle $ (here $p({\bf
x})=\vert q({\bf x})\vert - \langle \vert q \vert
\rangle $, and $\langle .. \rangle$ denotes averaging over the whole
system). We define the length scales $L(t)$, $L_q(t)$ and $L_p(t)$ as
the half-widths of the central maxima of the correlation functions
$C(r,t)$, $C_q(r,t)$ and $C_p(r,t)$, respectively. We find that these
length scales grow differently from each other (see the inset in
Fig.~(\ref{correl})) and from the average separation of topological
objects $D(t)$, indicating that dynamical scaling is
violated. The half-widths $L_q(t)$ and $L_p(t)$ do not grow as
power-laws of time, but rather as $a \log(b t)$, where a and b are
constants. The half-width of the spin-spin correlation, $L(t)$, grows
much faster, and at late times fits the power law
$L(t) \propto t^{0.38 \pm 0.02}$ \cite{Toyoki-note}.

We furthermore find that each family of the correlation functions
individually does not collapse onto a universal curve, providing a
further indication of the violation of dynamical scaling.  In
Fig.~(\ref{correl}), we show an attempt to collapse the correlation
functions $C(r,t)$, $C_q(r,t)$ and $C_p(r,t)$ using the lengths
$L(t)$, $L_q(t)$ and $L_p(t)$. The lack of collapse is most readily
apparent in the topological charge correlation $C_q(r,t)$ \cite{APS}.
A more detailed discussion of our results for the topological
correlation functions will be given separately.

In conclusion, we have characterised the phase ordering kinetics in
the investigated system in terms of the distinct processes of single
texture unwindings and topological charge annihilation, and
demonstrated that dynamical scaling is strongly violated.
Most of the methods developed in this paper should be equally well applicable
to phase ordering in systems with topological textures in higher
dimensions.


\acknowledgements
We thank J.~Borrill, A.~J.~Bray, N.~Goldenfeld, and Y.~Oono for useful
discussions, and the Isaac Newton Institute, where this collaboration
was started, for hospitality. M.~Z. acknowledges support by the
U.~S.~National Science Foundation through Grant NSF-DMR-89-20538.


\begin{figure}
\caption{Surface plots of the topological charge density $q({\bf r})$
in a $85 \times 85$ section of a phase ordering system
at time (a) t=0.0356, (b) t=0.126, and (c) t=1.59. The horizontal
scale is in units of one lattice spacing.
\label{big}}
\end{figure}
\begin{figure}
\caption{The process of topological charge annihilation, starting from
an overlapping full texture and antitexture. The plots show $q({\bf
r})$ for (a) $t=0$ (initial configuration) and (b) $t=3.0$. Note that
the vertical scale is different in (a) and (b). The quantity $P=\int
d^2r |q({\bf r}|$ decays from 1.54 in (a) to 0.89 in (b).
\label{annih}}
\end{figure}
\begin{figure}
\caption{Time dependences of the quantities $Q_-(t)$ (lower curve),
$Q(t)$ (middle curve), and $Q_+(t)$ (upper curve) during phase
ordering of a system of size $252 \times 252$.
 Inset: Time dependence of $q_{max}(t)$.
\label{steps}}
\end{figure}
\begin{figure}
\caption{Time dependences of the integrated absolute charge $P(t)$ and
the free energy $E(t)=\int d^2r\,{1 \over 8 \pi} ({\bf \nabla m})^2$
during phase ordering of a system of size $512 \times 512$, averaged
over 14 initial conditions. The dashed line has a slope of $-0.64$.
Inset: The decay of the difference
$E(t)-P(t)$. The full line has a slope of $-0.92$.
\label{energy}}
\end{figure}
\begin{figure}
\caption{The rescaled correlation functions $C(r/L(t))$, $C_q(r/L_q(t))$ and
$C_p(r/L_p(t))$ at specified times (see the main text for definitions). Inset:
growth of
the length scales $L(t)$, $L_q(t)$ and $L_p(t)$ (note that the graph is
semilogarithmic).
\label{correl}}
\end{figure}



\begin{references}
\bibitem{annihilation} Details of our investigation of the
annihilation process will be reported separately.
\bibitem{ineq} Neighboring regions with small amounts of positive and
negative topological charge, undergoing annihilation, can have $E$
substantially larger than $P$ [10]. However, the contribution of such
regions to the total values of $E$ and $P$ in the whole system is small.
\bibitem{Toyoki-note} This is roughly
consistent with the result $L(t) \propto t^{0.42 \pm 0.03}$ in [4].
\end{references}
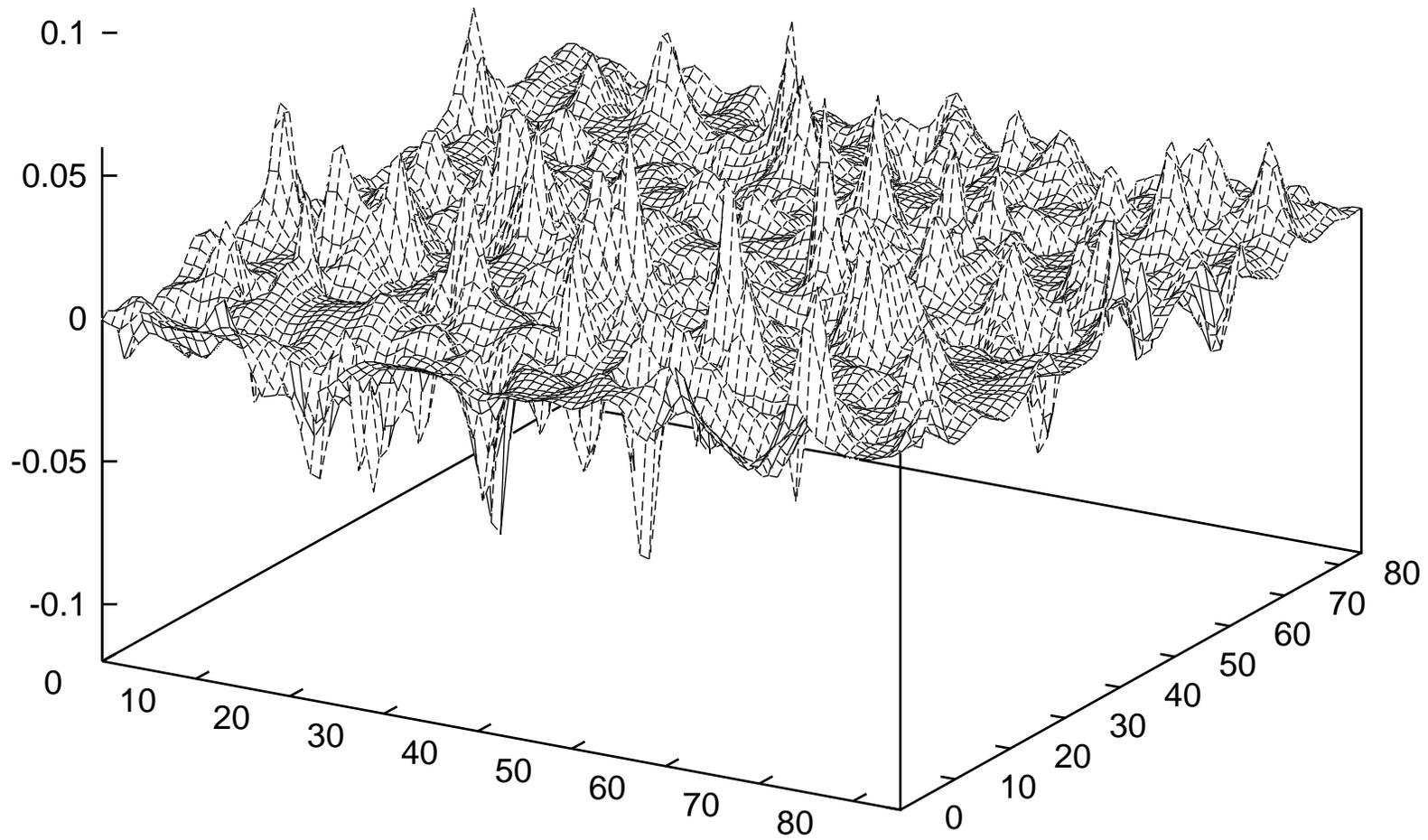
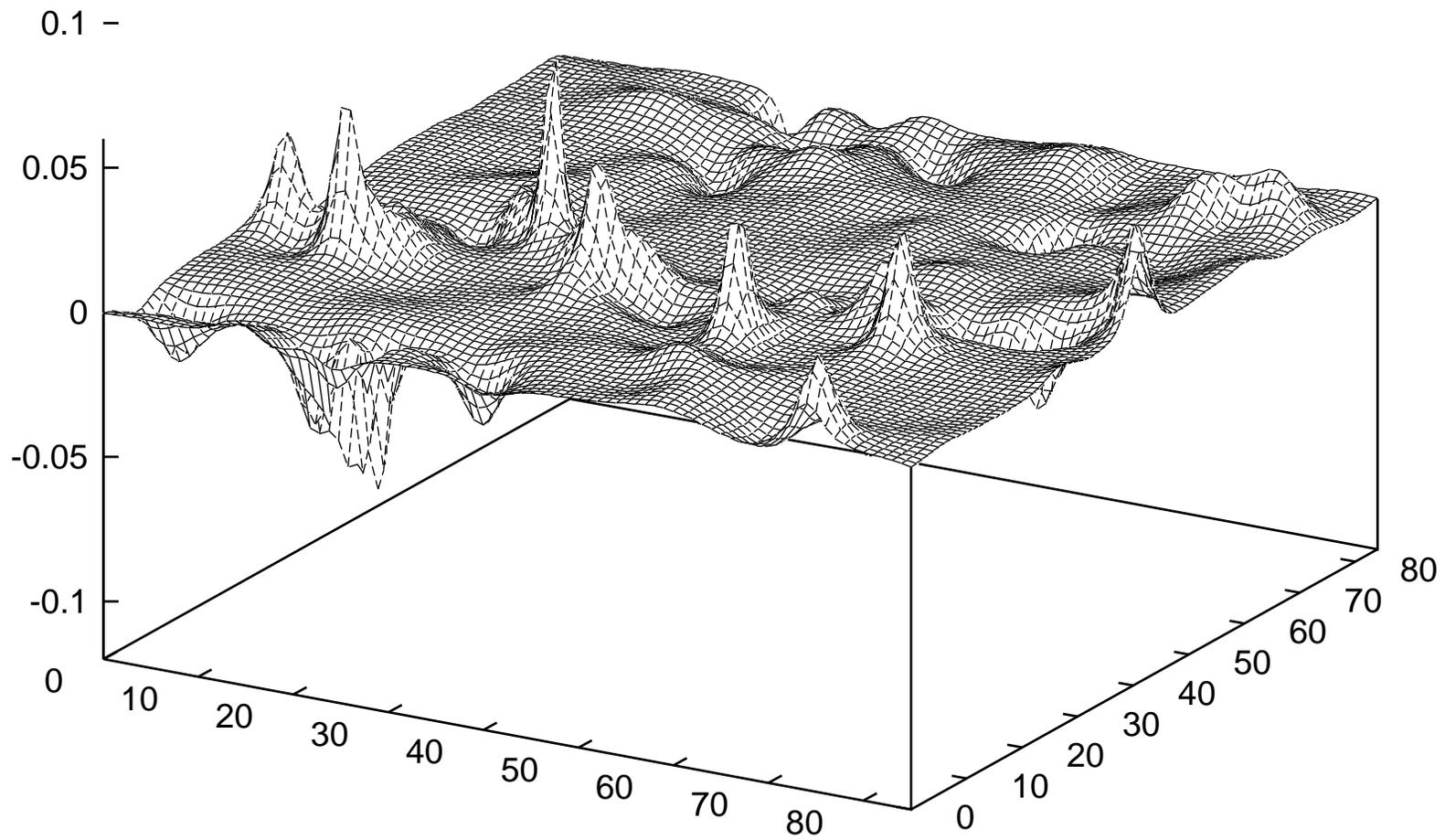
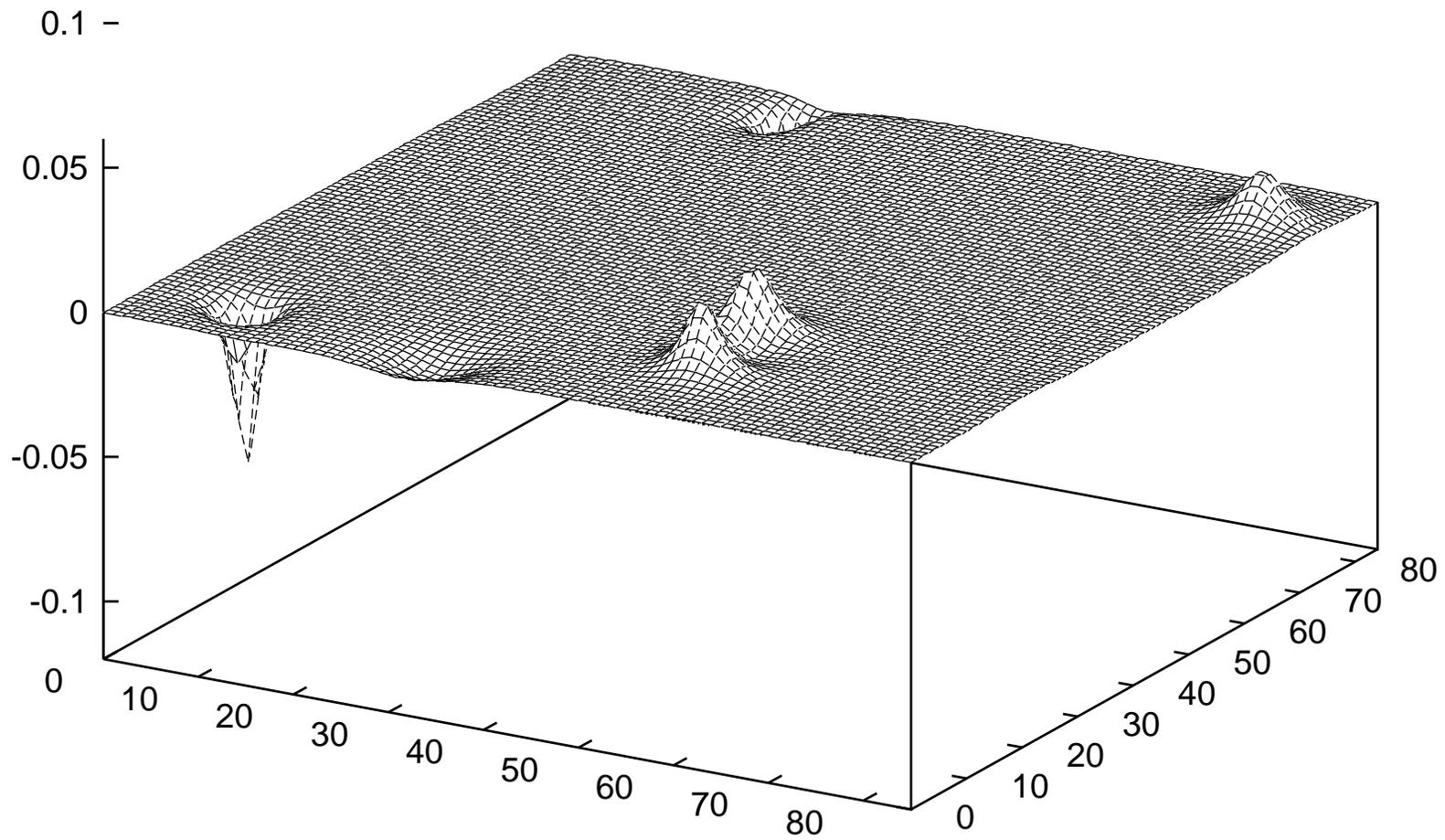
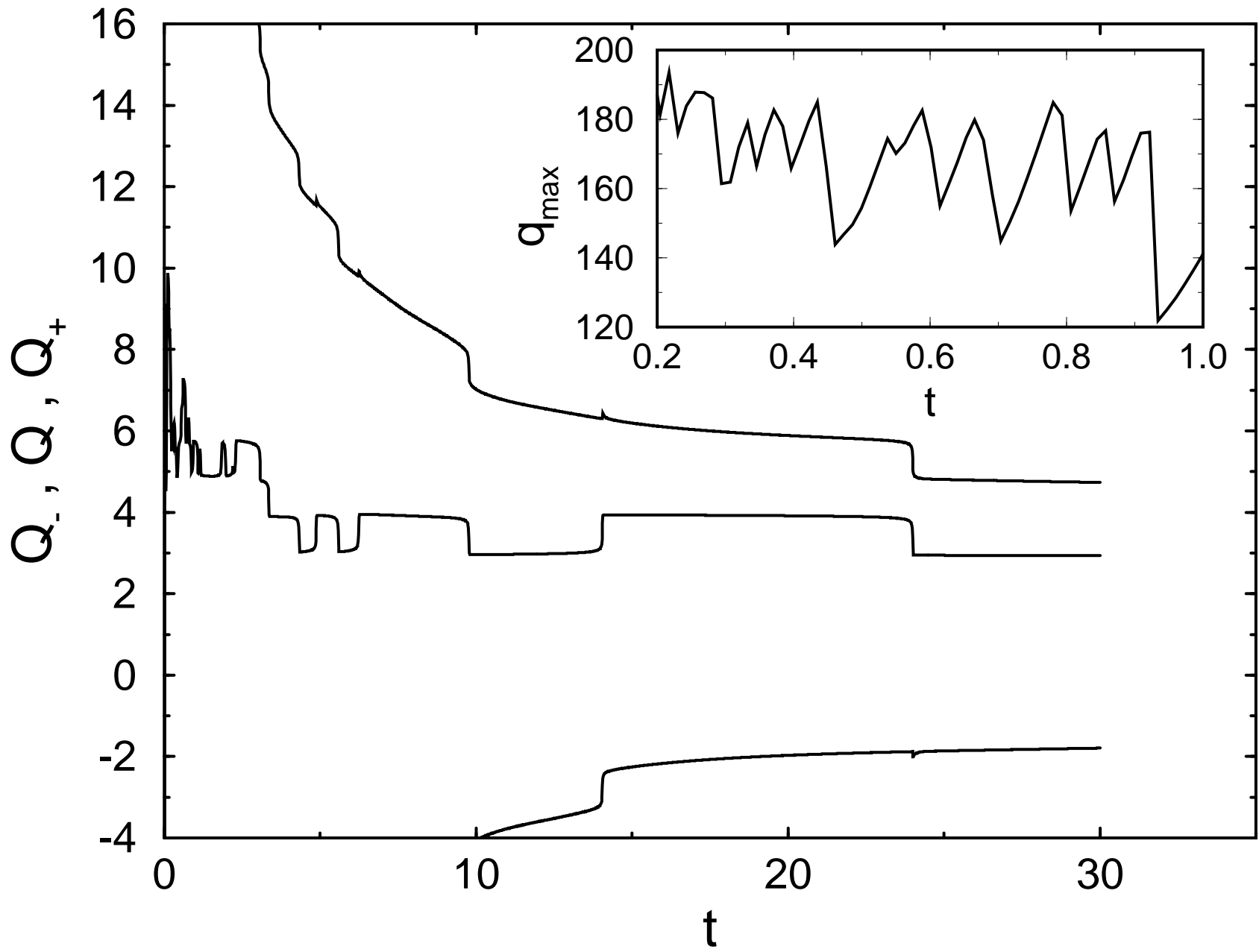
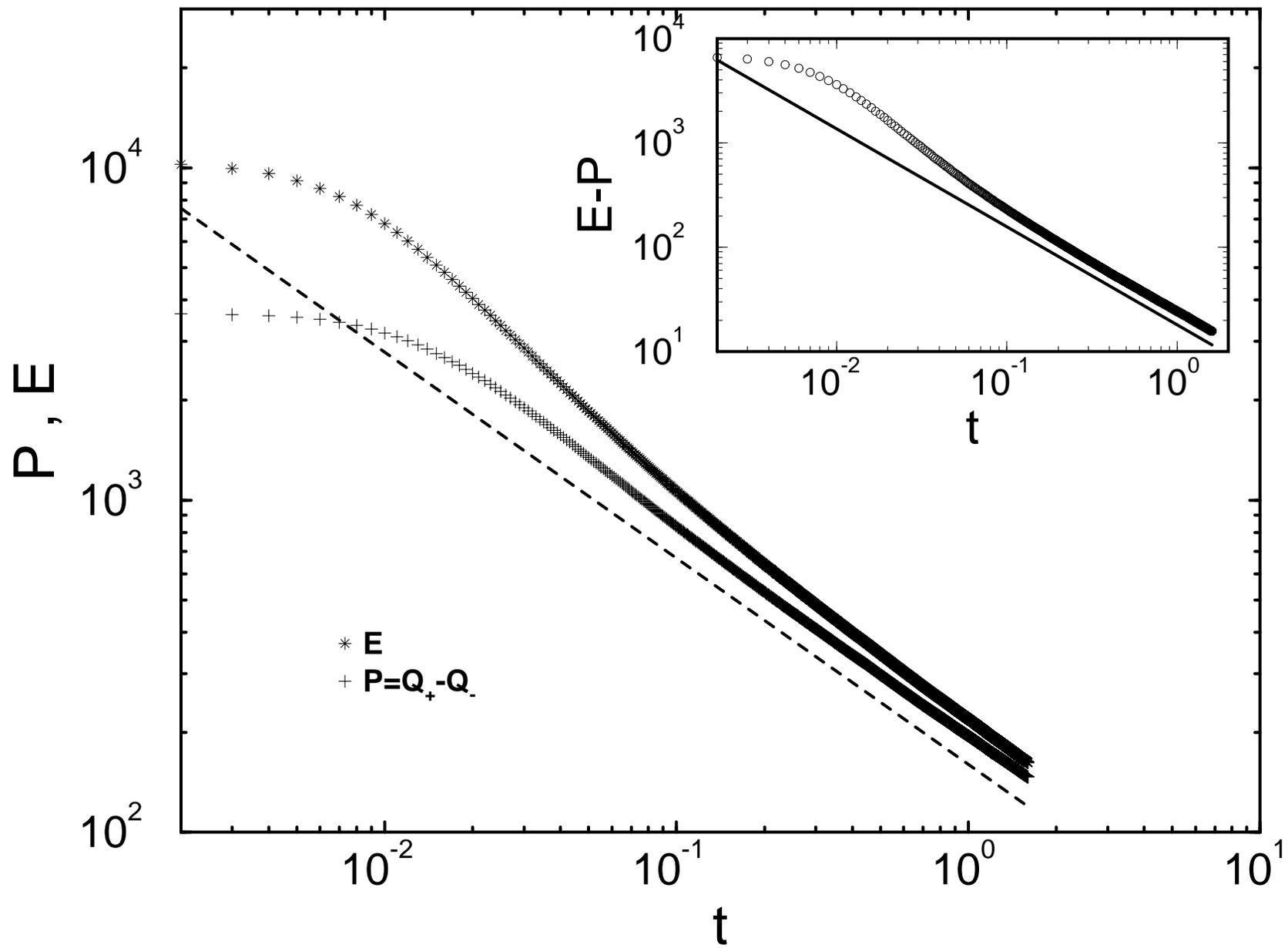
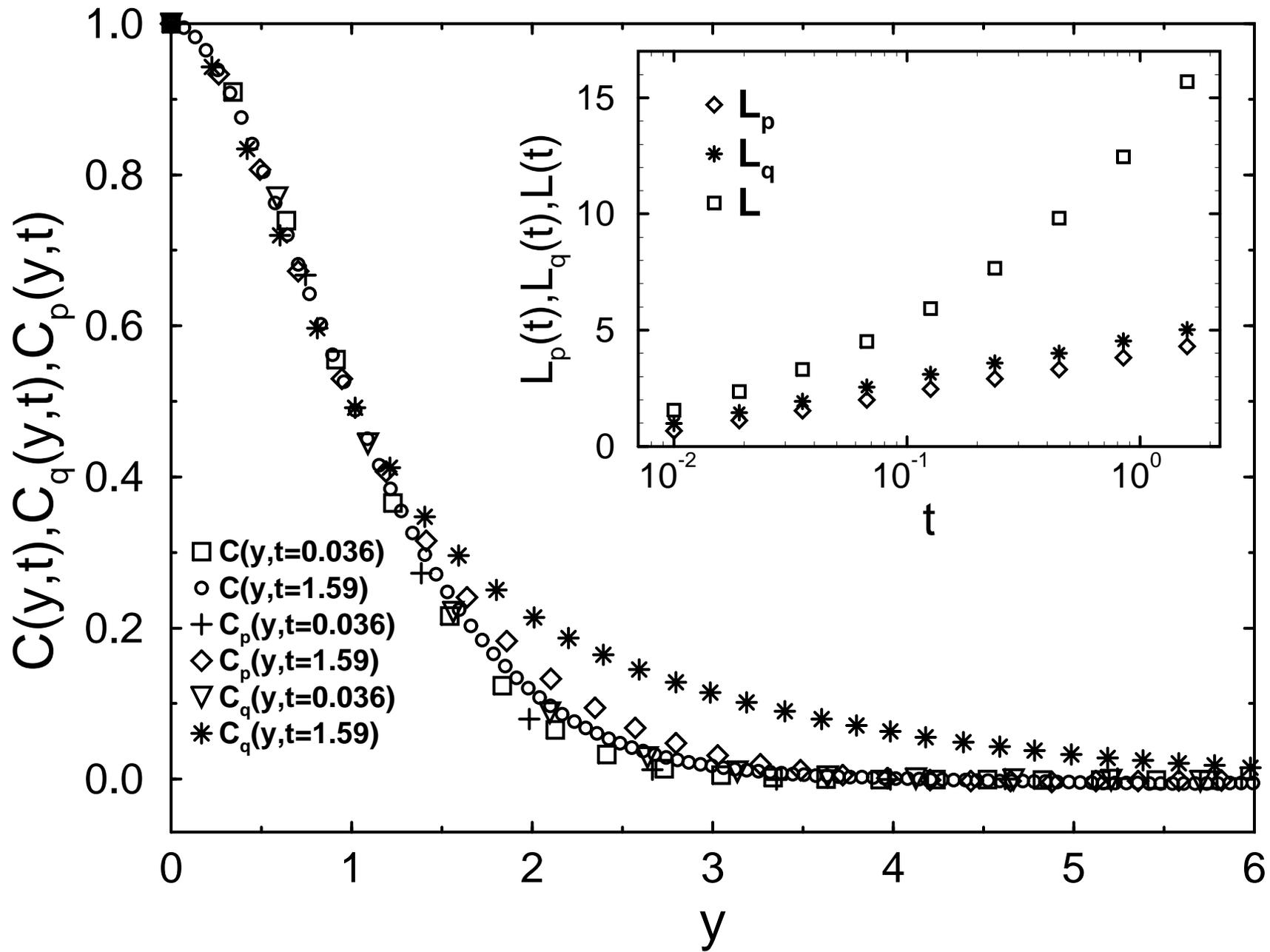
\end{document}